# Optical generation and detection of local non-equilibrium phonons in suspended graphene

Sean Sullivan,[1] Ajit Vallabhaneni,[2] Iskandar Kholmanov,[3] Xiulin Ruan,[2] Jayathi Murthy,[4] Li Shi[1,3*]

[1]*Materials Science and Engineering Program,* [3]*Department of Mechanical Engineering, University of Texas at Austin, Austin, Texas 78712*
[2]*School of Mechanical Engineering, Purdue University, West Lafayette, Indiana 47907*
[4]*Department of Mechanical and Aerospace Engineering, University of California, Los Angeles, California 90095*

*Email: lishi@mail.utexas.edu

**Abstract**

The measured frequencies and intensities of different first- and second- order Raman peaks of suspended graphene are used to show that optical phonons and different acoustic phonon polarizations are driven out of local equilibrium inside a sub-micron laser spot. The experimental results are correlated with a first principles-based multiple temperature model to suggest a considerably lower equivalent local temperature of the flexural phonons than those of other phonon polarizations. The finding reveals weak coupling between the flexural modes with hot electrons and optical phonons. Since the ultrahigh intrinsic thermal conductivity of graphene has been largely attributed to contributions from the flexural phonons, the observed local non-equilibrium phenomena have important implications for understanding energy dissipation processes in graphene-based electronic and optoelectronic devices, as well as in Raman measurements of thermal transport in graphene and other two-dimensional materials.

The superior electronic and thermal properties of graphene have been subjects of active research because they may enable the use of graphene in electronic and optoelectronic devices,[1–4] and also as a thermal management material.[5–8] The performance of functional graphene devices is intimately coupled to the scattering processes between the electrons and phonons, as well as scattering between phonons of different polarizations. Upon optical or electrical excitation, hot charge carriers in graphene with energies of hundreds of meV scatter with one another to establish local thermal equilibrium among themselves within a time scale and length scale on the order of 0.1 ps and 100 nm, respectively, according to several theoretical and experimental studies.[9–12] These charge carriers are scattered with optical and acoustic phonons in graphene at different time or length scales depending on the electron energy, phonon modes, defects, and interactions with the environment. Emission of optical phonons plays a dominant role in the relaxation of hot carriers with energies above about 196 meV,[13] which is comparable to the energy of in-plane polarized optical phonons in graphene. In comparison, charge carriers with energies well below 196 meV rely on scattering with acoustic phonons to relax their energy to the lattice.[13] In defect-free graphene, scattering between charge carriers and acoustic phonons with a wavevector larger than the Fermi wavevector is restricted due to the stringent momentum conservation requirement. The resulting low scattering rate of charge carriers by acoustic phonons has been used to explain the ultrahigh electron mobility found for suspended clean graphene under low electric bias,[14] as well as the enhanced photoresponse by carrier multiplication in graphene photodetectors.[15,16] On the other hand, disorder and defects in graphene can relax the momentum conservation requirement and facilitate scattering of charge carriers with the entire thermal phonon distribution via supercollisions, even when the phonon temperature is above the Bloch-Grüneisen temperature, $T_{BG}$, and acoustic phonons with a wavevector larger than the Fermi wavevector in graphene are thermally populated.[17,18] In addition, charge carriers in supported graphene are also scattered by surface polar optical phonons and charged impurities in the substrate, which considerably suppresses the electron mobility in supported graphene.[19]

Despite the extrinsic effects due to disorder, substrate support, and surface contamination, several ultrafast optical pump-probe measurements of supported and suspended graphene samples have suggested rapid relaxation between the hot carriers and optical phonons within a time scale of 50-150 fs,[12,20–23] and slow relaxation of the resulting hot optical phonons with acoustic phonons over a time scale of 2 to 3 ps.[12,21] These experiments have directly obtained the time scales for the hot electrons or optical phonons to cool back to the ambient temperature, while the acoustic phonon temperatures were not measured and were assumed to be much lower than the hot electron and optical phonon temperatures. However, considerable heating of the acoustic phonons were found in measurements of high-field graphene electronic devices,[24–26] although the acoustic phonon temperature rises were likely lower than the optical phonon temperature, which was found to be close to the hot electron temperature.[24,25] If the temperature rise of the acoustic phonons was not negligible in the pump-probe measurements, one would expect the observed cooling time scales to be longer than the relaxation time scales between the hot energy carriers and the acoustic phonons, because it would take additional time for the acoustic phonons to cool back to the ambient temperature after they were thermalized with the hot electrons or optical phonons. In addition, it remains to be seen whether or not local equilibrium has been established among different acoustic phonon polarizations in either photoexcited or electrically biased graphene.

Clarifying this question is not only important for establishing a better understanding of the energy carrier transport processes in graphene-based devices, but also for the correct interpretation of optical measurements of thermal transport in graphene. In particular, micro-Raman thermometry techniques have become a popular approach for thermal transport measurements of graphene[27–30] and other two-dimensional (2D) materials.[31,32] These measurements rely on the optical heating of the sample by a focused laser beam, and have yielded a wide range of thermal conductivity values for graphene. The variation has been partly attributed to the very different optical absorption values used in the data analysis of different measurements.[27,29,30] However, it is unclear whether the varying thermal conductivity results can also be caused by the use of different features in the Raman spectra, including the peak position and intensity, to extract the graphene temperature. In one Raman measurement,[33] the thermalization length between electrons and phonons is taken to be about 1 μm based on prior theoretical and experimental reports.[13,23] This length scale, instead of the 0.5-1 μm laser beam spot size, is taken to be the effective hot spot size in the numerical solution of the heat diffusion equation. In another prior Raman measurement,[27] the observed thermal resistance of suspended graphene at different laser spot sizes was used to suggest that low-frequency acoustic phonons were in the quasi-ballistic transport regime, took a lower temperature than higher frequency phonons, and carried less heat than in the diffusive regime. Moreover, a recent first principles-based multi-temperature model has predicted that electrons and different phonon modes are driven out of local thermal equilibrium under laser excitation.[34] However, there is a lack of direct experimental evidence regarding whether local non-equilibrium is sustained between different phonon populations in the graphene sample during micro-Raman measurements. Such local non-equilibrium not only influences the temperature measurement from the Raman spectra, but would also reduce the thermal conductivity contribution from the low-frequency phonons, which are predicted to dominate the thermal transport in clean suspended graphene when the different phonon polarizations are not driven out of local equilibrium.[35,36] In both transport measurements and device applications of graphene, the degree of local non-equilibrium between different acoustic and optical phonon polarizations is a fundamental property that requires further investigation.

Here, we report an optical experiment for driving and observing local non-equilibrium between optical and different acoustic phonon polarizations in suspended graphene within a submicron laser spot. While the optical phonon temperature is obtained from the measured Raman peak intensities as in prior works,[24,25,29,37] we use the positions of not only the first-order Raman peak but also three second-order Raman peaks to find that the equivalent local temperatures of different acoustic phonon polarizations are lower than the optical phonon temperature. According to a multi-temperature model, furthermore, the experimental results reflect that local non-equilibrium exists among different acoustic phonon polarizations, with the out-of-plane polarized flexural (ZA) modes especially under-populated in comparison with the in-plane polarized transverse and longitudinal acoustic (TA and LA) modes. Based on the observed local non-equilibrium between optical phonons and different acoustic phonon polarizations, the prior descriptions of the energy dissipation mechanisms in optically and electrically excited graphene devices should be revised.

In our experiments, we used micro-Raman spectroscopy to probe the local temperatures of different phonon populations in suspended graphene optically heated by the focused probe laser beam. The graphene samples were grown on copper foils using low-pressure chemical vapor deposition (LPCVD), and subsequently transferred to 400 nm-thick low-stress silicon nitride membranes with arrays of holes ranging in diameter from 5 to 20 μm, which were fabricated with focused ion beam milling, as shown in Fig. 1. During the Raman measurements, the graphene sample was placed inside a vacuum chamber evacuated to a pressure of ~1 torr with a 150 μm-thick cover glass slide above and a heating stage below the graphene sample. A 532 nm-wavelength laser beam was focused by a 50x achromatic objective lens in a backscattering geometry through the cover glass at the center of the suspended graphene. The radius of the Gaussian laser beam focused through the cover glass was measured to be 360 ± 3 nm based on the Raman intensity profile obtained across a cleaved Si edge at all laser powers used.[27]

The Raman spectra of the suspended graphene sample were obtained at different laser powers and stage temperatures. Figure 1 shows a representative Raman spectrum that contains four Stokes peaks, including the pronounced G-band and 2D-band, which are found near 1580 cm$^{-1}$ and 2670 cm$^{-1}$, respectively. The absence of the D peak at around 1350 cm$^{-1}$, which is associated with defects, suggests the high quality of the graphene sample.[38,39] The Stokes and anti-Stokes G-band in graphene are associated with the emission and absorption of a zone-center longitudinal optical (LO) phonon by an excitation photon. In comparison, the 2D-band involves the absorption or emission of two in-plane transverse optical (TO) phonons with opposite and relatively large wavevectors in second-order, double resonant processes.[40–42] Besides the G and 2D peaks, which have been commonly used for transport measurements of graphene, we examine the details of two other second-order Raman peaks, as labeled in Fig. 1. The first one is the D+D'' peak near 2450 cm$^{-1}$, which is attributed mainly to the scattering of the electronic excitation with both the aforementioned TO phonon and a longitudinal acoustic (LA) phonon along the Γ−K direction.[38,43] The second one is the 2D' peak near 3240 cm$^{-1}$, which involves two LO phonons with small, opposite wavevectors near the zone-center.[38] The frequency of the LA (D'') phonon near 1120 cm$^{-1}$ can be obtained as $\omega_{D''} = \omega_{(D+D'')} - \omega_{2D}/2$ as discussed in prior works.[38,39,41] As the stage temperature or laser power is increased, the peak frequencies of the four Raman bands downshift, while the anti-Stokes G-band intensity ($I_{AS}$) increases relative to the Stokes peak intensity ($I_S$), as shown in Figs. 1, 2, and 3.

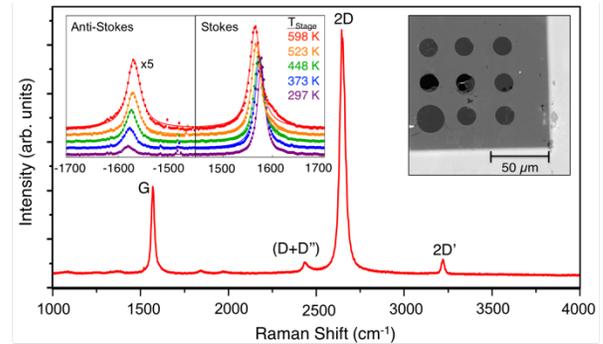

**Figure 1:** Representative Raman spectra of the suspended graphene sample taken at the highest laser power (4.7 mW), with the stage temperature kept at 598 K. The top left inset shows the anti-Stokes and Stokes G-band obtained at 4.7 mW incident laser power and different stage temperatures. The top right inset shows a scanning electron micrograph (SEM) of the graphene sample on a holey silicon nitride membrane.

The Stokes and anti-Stokes intensities of the G-mode are proportional to $n_G+1$ and $n_G$, respectively, where $n_G$ is the occupation of the zone-center optical phonon mode. The absolute intensities are affected by the measurement setup and the Raman cross-section of the corresponding phonon mode. However, taking the ratio of the anti-Stokes to Stokes intensity cancels many mode-specific terms, leaving

$$\frac{I_{AS}}{I_S} = C \left(\frac{\omega_L+\omega_G}{\omega_L-\omega_G}\right)^4 \exp\left(\frac{-\hbar\omega_G}{k_B T_O}\right) \quad (1)$$

in which $C$ is a constant that depends on the optical collection efficiencies of the Stokes and anti-Stokes peaks and can take into account the slight difference between the anti-Stokes and Stokes phonons,[24,38] $\omega_L$ and $\omega_G$ are the laser and the measured zone-center optical phonon frequencies, respectively, and $T_O$ is the equivalent temperature that can be used in the Bose-Einstein distribution to obtain the actual zone-center optical phonon population. Thus, the intensity ratio can be used to obtain $T_O$ provided that the constant $C$ can be calibrated for this measurement.

The laser power- and temperature-dependent anti-Stokes and Stokes intensity ratios of the G-band are shown in Fig. 2. For the highest stage temperature of 598 K, we extrapolate the intensity ratio as function of incident laser power to zero laser heating using a second order polynomial fit. With the optical phonon temperature equal to the stage temperature at zero laser power, the extrapolated intensity ratio at zero laser power is used in Equation (1) to obtain a calibration coefficient $C = 0.68 \pm 0.06$. The obtained $C$ range is subsequently used to obtain the corresponding $T_O$ at each laser power and stage temperature based on the measured intensity ratio and Equation (1). Due to the high optical phonon energy in graphene, the anti-Stokes peak is weak at the lowest laser power and lowest stage temperature conditions, for which the intensity ratio data cannot be obtained accurately to determine the optical phonon temperature.

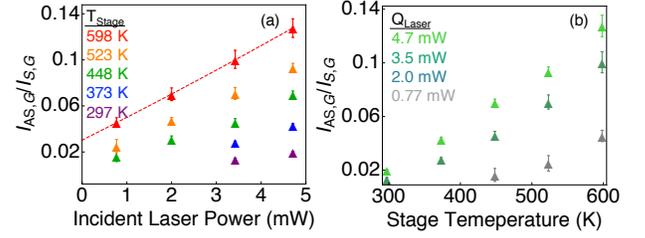

**Figure 2:** G-band integrated anti-Stokes/Stokes intensity ratios as a function of the incident laser power (a) and stage temperature (b). The dashed fitting line in (a) is used to determine the coefficient $C$ in Equation (1) by extrapolating the intensity ratio to zero laser heating.

In addition to the intensity ratio, the frequency shifts of different Raman peaks also contain information on the populations and equivalent temperatures of the phonons involved in the anharmonic decay processes of the Raman-active phonon mode. The frequencies of the Raman-active optical phonons are influenced by their anharmonic interactions with acoustic phonons as well as thermal expansion.[44] The negative thermal expansion of graphene contributes to an increase of the Raman mode frequency with increasing temperature. However, this effect is small compared to the frequency downshift caused by increasing anharmonic interactions with increasing temperature. Thus, the observed frequency downshift of the different Raman peaks of graphene during laser heating reflects an increase in the populations of those interacting phonon modes, which do not necessarily follow the equilibrium Bose-Einstein distribution given by a single local temperature.

By extrapolating the laser power-dependent peak shift data taken at the lowest stage temperature to zero laser power using a second order polynomial fit, as shown in Fig. 3(a-c), we find the Raman shift ($\omega_0$) in the case of no laser heating at room temperature ($T_\infty$) where the graphene phonons would be at equilibrium. The measured peak frequency at the lowest laser heating power downshifts nearly linearly with increasing stage temperature up to 598 K, as shown in Fig. 3(d-f). A linear fitting of the data obtained at the lowest laser power is used to obtain the temperature ($T$) coefficient of the peak position ($\omega$), $d\omega/dT$, for each peak. By taking $T^{eq} = (\omega - \omega_0)(d\omega/dT)^{-1} + T_{Stage}$, we determine the equivalent equilibrium temperature ($T^{eq}$) that would result in the same frequency shift ($\omega$) of each Raman peak as that measured at different laser powers. The obtained $T^{eq}$ reflects the populations and an average temperature of those phonons that are anharmonically coupled with the corresponding Raman-active phonons. The obtained peak frequencies for several measurements at high laser power and stage temperature values fall outside the range of the linear fitting line, and are not used for the temperature conversion.

$T_O$ is increasingly more pronounced when the peak shift used to determine the $T^{eq}$ changes from the G, 2D, 2D', (D+D''), to D'' frequency. At the highest laser power of 4.71 ± 0.08 mW and room temperature for the stage, we obtain $T_O$, $T^{eq}_G$, $T^{eq}_{2D}$, and $T^{eq}_{D''}$ of 536 ± 19 K, 498 ± 13 K, 460 ± 10 K, and 397 ± 20 K, respectively, based on the measured intensity ratio and the corresponding peak shifts.

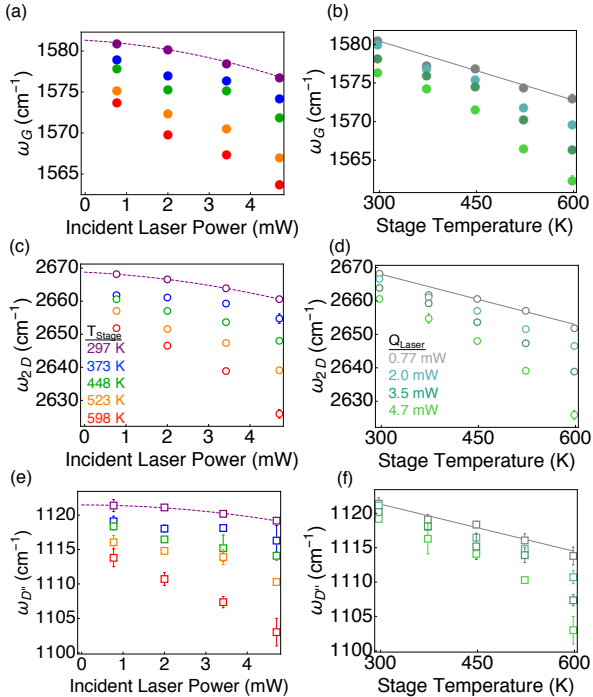

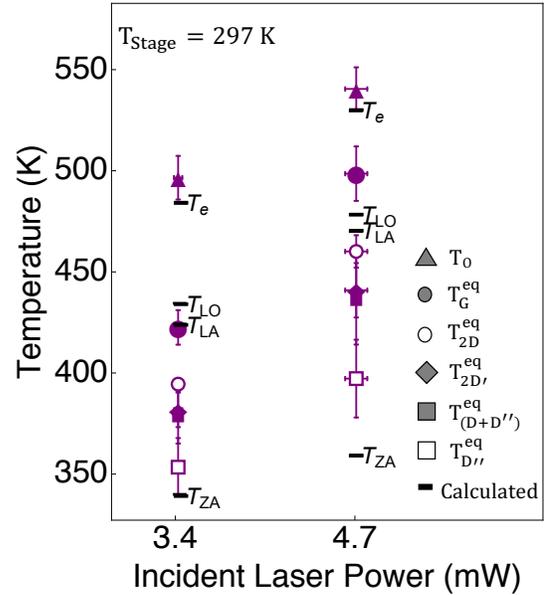

**Figure 3:** Stage heating and laser heating effects on the Raman spectra of suspended graphene. The peak shift as a function of laser heating (left column) and stage temperature (right column) for the G-band (a, b), 2D-band (c, d), and the D'' frequency (e, f). Dashed purple lines are second order polynomial fits for data taken at the lowest stage temperature as a function of the laser power. The solid gray lines are linear fits taken at the lowest laser power. Each data point represents the average and random uncertainty of between 6 and 27 measurements with double-sided 95% confidence. The legend in (c) is applicable to (a) and (e), while the legend in (d) is applicable to (b) and (f).

As shown in Fig. 4, the obtained $T^{eq}$ based on the peak shifts can be lower than the optical phonon temperature obtained from the intensity ratio. At the two highest laser powers, the deviation of $T^{eq}$ from

**Figure 4:** Measured phonon temperatures in the suspended graphene at the two highest incident laser powers and the lowest stage temperature (297 K). Triangles represent the optical phonon temperature values ($T_O$) extracted from the G-band anti-Stokes/Stokes intensity ratio, while the filled and open circles, filled diamonds, and filled and open squares are the equivalent temperature values measured from the peak shift of the G-band, 2D-band, 2D'-band, (D+D'')-band and D'' frequencies, respectively. The black horizontal lines indicate the calculated temperatures of the electrons ($T_e$), in-plane longitudinal optical and acoustic phonons ($T_{LO}$ and $T_{LA}$, respectively), and out-of-plane flexural phonons ($T_{ZA}$) for suspended graphene with an electronic thermal conductivity of $\kappa_e = 20 T_e / 300 \, K \, Wm^{-1}K^{-1}$.

As a comparison, Figure 4 shows the results of a first principles-based multi-temperature model of the Raman measurement, where the different relaxation times have been calculated from density functional perturbation theory.[34] At 4.71 mW laser power and 297 K stage temperature, the calculation has obtained $T_e$ = 530 K, $T_{LO}$ = 478 K, and $T_{LA}$ = 470 K, $T_{ZA}$ = 359 K for the local temperatures of electrons, LO phonons, LA phonons, and ZA phonons inside the Raman laser spot. The 110 K difference between the calculated $T_{LA}$ and $T_{ZA}$ is considerably larger than the 52 K difference between $T_e$ and $T_{LO}$, which is in turn larger than the 8 K difference between $T_{LO}$ and $T_{LA}$. The calculation result suggests that under these conditions, the largest non-equilibrium is actually between the ZA phonons and the in-plane polarized LA and TA phonons, rather than between the hot electrons and optical phonons or between optical phonons and LA or TA phonons. The underlying mechanism is the restrictive selection rule on the scattering of ZA phonons with either electrons or other phonon polarizations. Due to the reflection symmetry of monolayer suspended graphene, such scattering processes cannot involve an odd number of ZA phonons.[36,45,46]

The key feature of this theoretical finding is revealed by the experimental results. The obtained $T_O$ from the intensity ratio of the G-band (LO phonon) is somewhat higher than the calculated $T_{LO}$, likely due to the ignorance of defects in the calculation since defect scattering can potentially reduce the thermal conductivity contributions from different phonon polarizations. Meanwhile, the measured $T^{eq}_G$ is comparable to the calculated $T_{LA}$ and $T_{TA}$ values. The G-mode LO phonon near the zone center is mainly scattered with two intermediate frequency phonons (IFPs) of equal and opposite wave vectors and in the LA and TA branches,[44] so that the $T^{eq}_G$ is expected to be dominated by the population of LA and TA phonons away from the zone center. In comparison, the 2D-peak arises from Raman scattering of TO phonons near the K-point. In addition, the TO phonon of the 2D-peak can decay into another optical phonon and a low-frequency acoustic phonon near the zone center in addition to scattering with two IFPs.[38,41,42,44] Thus, $T^{eq}_{2D}$ is more sensitive to the population or temperature of the low-frequency acoustic phonons than the $T^{eq}_G$ determined from the G-peak shift. The measurement of $T^{eq}_{2D}$ lower than $T^{eq}_G$ therefore suggests a lower temperature of the low-frequency acoustic phonons near the zone center than the temperature of intermediate-frequency TA and LA phonons, which is reflected by $T^{eq}_G$. In comparison, the 2D'-band arises from electron scattering with two LO phonons with equal and opposite wavevectors. This scattering occurs within a single K-valley, as opposed to the intervalley process that gives rise to the 2D band, before interacting with IFPs that ultimately decay into TA and LA phonons.[38] Despite this fine distinction, the obtained $T^{eq}_{2D'}$ is comparable to $T^{eq}_{2D}$ because the decay processes involve similar types of acoustic phonons. In comparison, the (D+D")-peak results from the scattering of the electronic excitation with one TO phonon (D phonon) and a LA (D" phonon) near the K-point. Since the D" phonon can scatter with a pair of ZA phonons, the D" frequency is affected by the population of ZA phonons. Hence, the $T^{eq}_{D"}$ measured from the D" peak is directly influenced by the local temperature of the ZA phonons. Given the interactions between the D" phonons and the ZA phonons, the much lower measured $T^{eq}_{D"}$ compared to other measured temperatures is in qualitative agreement with the theoretical prediction of the lowest ZA temperature in the laser spot. In principle, local non-equilibrium between different phonon polarizations can be expected when the laser spot size is smaller than the corresponding relaxation or thermalization length. The observed non-equilibrium suggests that the thermalization lengths between ZA and other acoustic and optical phonon polarizations are larger than the submicron laser spot size, and should be on the micron scale.

The non-equilibrium is seen most clearly at the lowest stage temperature and the two highest laser power values, and can also be observed at 2 mW laser power and 523 K stage temperature. At the lowest incident laser power of 0.77 mW and the stage temperatures of 448 K and above, the differences in the temperature values obtained from the peak shifts and intensity ratio are within the appreciable uncertainty in the measured temperatures. In addition, the anti-Stokes intensity became too weak to be measured accurately when the stage temperature was reduced below 448 K at a laser power below about 2 mW, or the laser power is reduced below 0.77 mW even at the highest stage temperature of 598 K. Although the anti-Stokes peak measured at the lowest laser power can be increased by increasing the stage temperature, the suspended graphene samples could be damaged when the stage temperature was increased above 600 K.

While the Raman measurement results are able to reveal the local non-equilibrium between different phonon polarizations only at the relatively high incident laser power due to the limited temperature sensitivity of the Raman thermometry techniques, a recent multiple-temperature model calculation has shown that non-equilibrium is generated at laser powers about two orders lower than those used in our experiments.[34] The degree of local non-equilibrium can be revealed by the ratios between the ZA phonon temperature rise and the temperature rise of other energy carriers. These ratios calculated with the multi-temperature model for our experiments actually decrease with either decreasing laser power or decreasing stage temperature, suggesting more pronounced non-equilibrium at a lower laser power or lower stage temperature for the power and temperature ranges of our measurements. Although local non-equilibrium is expected to vanish at zero laser power, these results suggest that the degree of local non-equilibrium does not decrease monotonically with decreasing laser power. The underlying mechanism is that the relaxation length between the electronic excitations and different phonon polarizations increases with a decreasing phonon population when either the stage temperature or the laser power is decreased because of decreasing rates of electron-phonon and phonon-phonon scattering processes. Since local non-equilibrium between different energy carriers is measured when the corresponding scattering length is larger than the laser spot, these ratios of the calculated temperature rises decrease as the laser power or stage temperature is lowered and the scattering length is increased, at least for the laser power range investigated in the current work as well as in the prior multi-temperature model calculation for laser powers as low as 0.01 mW. [34]

In addition, it is necessary to point out that the electron relaxation time measured in a prior THz pump-probe measurement[20] is the time scale for the electrons to cool to the ambient temperature. Although the thermalization time between the electrons and different phonon populations are expected to increase with decreasing laser power or electrical field due to a reduced phonon population, the electron cooling time observed in the THz pump-probe measurement can still increase with increasing electric field, as reported,[20] because it takes additional time for the hot acoustic and ZA phonons excited at a high electric field to cool to the ambient temperature after the local thermal equilibrium is established between electrons and phonons. Since both the measurements and the multi-temperature model reported here reveal considerable heating of the acoustic phonons, especially the LA and TA phonons, it is necessary to measure the local temperatures of electrons and different phonon polarizations in order to obtain the thermalization length or time instead of just the electron or optical phonon cooling time scales. In addition, although the local ZA phonon temperature is much lower than the hot electron temperature, ZA phonons still carry a large fraction of the heat current according to the multi-temperature model,[34] because of the large specific heat capacity and thermal conductivity contribution of the ZA phonons at the temperature range, which is still low compared to the zone boundary frequencies of the other acoustic and optical phonon polarizations.

The observed non-equilibrium between different phonon polarizations, especially between the ZA and other phonons, has implications for both photo-excited and electrically biased graphene devices. The laser power used in the experiment yields an electric field on the order of 1 V/μm, which is comparable to those encountered in graphene electronic devices. In addition, the observed local non-equilibrium has been ignored in past heat diffusion analyses of Raman thermal transport measurements of graphene. To illustrate the effects of this assumption, we employ the analytic solution derived in a prior work to extract the thermal conductivity of suspended graphene.[27] Figure 5 displays this apparent thermal conductivity versus the apparent graphene temperature, as measured at the highest laser power. At a given stage temperature, the thermal conductivities extracted from a particular peak shift or from the anti-Stokes/Stokes ratio can vary widely. The apparent temperature values measured from the peak shift of Raman bands that involve lower frequency phonons, like the D+D" peak, are especially lower than those measured directly from the zone-center LO phonon population. These higher apparent temperatures result in lower apparent thermal conductivities. This feature can partly explain the relatively low thermal conductivity value reported in an earlier work based on the measured intensity ratio,[29] compared to those obtained based on the peak frequency shifts.[27,28,30,47] The values obtained here based on the peak shift are in the range of those reported from two prior Raman measurements of similar CVD graphene samples.[27,28] Thermal conductivity is inherently a diffusive property that can be obtained only when local thermal equilibrium is established. As such, the

Raman measurement cannot provide sufficient information to determine the exact thermal conductivity value of the graphene sample when the small laser spot size results in large local non-equilibrium among different phonons. Nevertheless, the difference between the intrinsic value obtained from first principle calculations and the obtained apparent thermal conductivity is comparable to the large uncertainty that is acknowledged in several prior Raman measurements.[27,28]

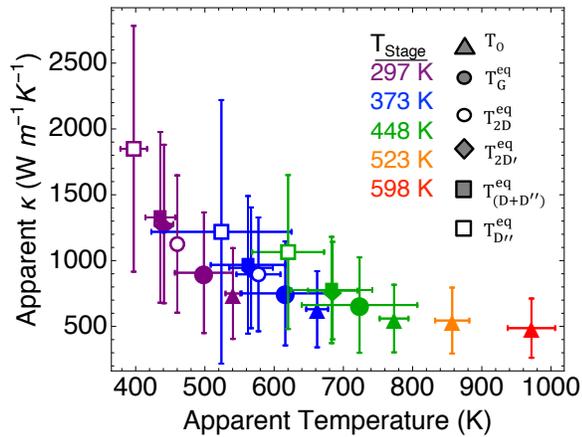

**Figure 5:** Apparent thermal conductivity value of the suspended graphene as a function of the apparent temperature measured at the highest laser power, 4.7 mW, using different phonon temperatures extracted from the Raman spectra, as indicted in the legend.

In tandem, the experimental and calculation results show that the out-of-plane polarized flexural phonons are underpopulated compared to not only the optical phonons but also other in-plane polarized acoustic phonons inside a submicron laser spot focused on suspended graphene. This finding is based on the examination of not only the intensity ratio and the first-order Raman peak position, but also three second-order Raman peak positions that contain information on the local temperatures of the acoustic phonons and, especially, the flexural phonons. The observed local non-equilibrium suggests that the thermalization length between the acoustic phonons – in particular, the flexural modes – and the other energy excitations is larger than the Gaussian beam radius of 360 nm. These findings suggest that it is important to consider non-equilibrium not only between the hot charge carriers and the acoustic phonons, but also among the different optical and acoustic phonon populations in graphene electronic and optoelectronic devices, and especially between the heat-carrying flexural phonons and other phonon polarizations. In addition, the observed non-equilibrium and thermalization length have practical implications in the interpretation of Raman thermal transport measurements of graphene and other two-dimensional materials with potentially long thermalization lengths between different energy excitations. Specifically, the shifts of different Raman peaks should be used to evaluate whether local equilibrium can be assumed in future Raman thermal transport measurements, if the intensity ratio of the anti-Stokes peak cannot be measured accurately to obtain the optical phonon temperature.

**Acknowledgement**
The authors thank David Choi for assistance in preparing the silicon nitride membranes, and Dr. Dhruv Singh for fruitful discussions. The work is supported in part by the US National Science Foundation Thermal Transport Processes Program award CBET-1336968 (SS and LS), Air Force Office of Scientific Research (AFOSR) MURI award FA9550-12-1-0037 (IK, XR, and LS), and Air Force Office of Scientific Research award FA9550-11-1-0057 (XR).